\documentclass[11pt, oneside]{article}   	% use "amsart" instead of "article" for AMSLaTeX form++++\at
\usepackage{geometry}                		% See geometry.pdf to learn the layout options. There are lots.
\usepackage{graphicx}				% Use pdf, png, jpg, or eps§ with pdflatex; use eps in DVI mode
\usepackage{amsmath}								% TeX will automatically convert eps --> pdf in pdflatex		
\usepackage{amssymb}

\title{Inclusive scattering  matrix and scattering of quasiparticles}
\author{Albert Schwarz}
\date{}							% Activate to display a given date or no date

\begin{document}
\author {  A. Schwarz\\ Department of Mathematics\\ 
University of 
California \\ Davis, CA 95616, USA,\\ schwarz @math.ucdavis.edu}
\newcommand{\ba}{{\mathbf a}}
\newcommand{\la}{{a_{\text{L}}}}
\newcommand{\ra}{{a_{\text{R}}}}
\newcommand{\ta}{\tilde{a}}
\newcommand{\cA}{{\mathcal A}}
\newcommand{\bA}{{\mathbf A}}
\newcommand{\bB}{{\mathbf B}}
\newcommand{\cB}{{\mathcal B}}
\newcommand{\bC}{{\mathbb C}}
\newcommand{\cC}{{\mathcal C}}
\newcommand{\cF}{{\mathcal F}}
\newcommand{\tF}{{\tilde{F}}}
\newcommand{\bH}{{\bf H}}
\newcommand{\cH}{{\mathcal H}}
\newcommand{\tH}{\tilde{H}}
\newcommand{\tJ}{\tilde{J}}
\newcommand{\cL}{{\mathcal L}}
\newcommand{\tl}{{\tilde{l}}}
\newcommand{\bp}{{\bf p}}
\newcommand{\bP}{{\bf P}}
\newcommand{\bx}{{\bf x}}
\newcommand{\bR}{{\mathbb R}}
\newcommand{\res}{\text{res}}
\newcommand{\tS}{\tilde{S}}
\newcommand{\cT}{{\mathcal T}}
\newcommand{\tw}{\tilde{w}}
\newcommand{\tz}{\tilde{z}}
\newcommand{\bZ}{{\mathbb Z}}
\newcommand{\bX}{{\mathbf X}}
\newcommand{\bra}{\langle}
\newcommand{\ket}{\rangle}
\newcommand{\bv}{{\bf v}}

\maketitle
\abstract { The quantum theory can be formulated in the language  of positive functionals on Weyl or Clifford algebra ($L$-functionals).  It is shown that  this language gives simple understanding of diagrams  of Keldysh formalism (that coincide in our case with the diagrams of thermo-field dynamics). The  matrix elements of the scattering matrix in the formalism of $L$-functionals are related to inclusive cross-sections, therefore we suggest  the name "inclusive scattering matrix" for this notion.
The inclusive scattering matrix can be expressed in terms of on-shell  values of generalized Green functions. This notion is necessary if we want to analyze collisions of quasiparticles.}


\section {Introduction}
In this paper we analyze the scattering of elementary excitations of equilibrium state or, more generally,  of stationary translation-invariant state. In zero temperature case these excitations can be identified with particles, for $T>0$ they are quasiparticles.  We are using the technique of $L$-functionals
(positive functionals on the Weyl algebra) suggested in \cite {SCH}; see also \cite {T}. This technique was rediscovered  later under the name of thermo-field dynamics  or TFD in a  different and less transparent form (see  \cite {CU} for the review of TFD and Keldysh formalism, \cite {K}  or \cite {MU} for review of  Keldysh formalism). We give a very simple derivation of the diagram technique for the calculation  of adiabatic  S-matrix and  generalized Green functions (GGreen functions) in the formalism of $L$-functionals  (this technique in the case at hand coincides with TFD and Keldysh techniques).   Scattering matrix in the formalism of $L$-functionals denoted $\hat S$ can be defined in terms of adiabatic S-matrix  (see (\ref {SSS})). Its  matrix elements are related to inclusive cross-sections; therefore we suggest the name "inclusive scattering matrix" for $\hat S$. We obtained  diagram techniques for calculation of $\hat S$, hence for calculation of inclusive cross-sections.   Recall that the inclusive cross-section of the process $(p_1, p_2\to q_1,..., q_n)$ is a sum (more precisely a sum of integrals) of cross-sections  of processes $(p_1, p_2\to q_1,...,q_n, r_1,....r_m)$ over all possible $(r_1,..., r_m).$ 

When we are talking about scattering of thermal quasiparticles we should restrict ourselves to stable (or, at least, almost stable) quasiparticles. However, it is unlikely that a collision of stable  quasiparticles will produce only stable quasiparticles; almost  always we have something else. Therefore the only reasonable notion in this case is the notion of inclusive cross-section. This is the reason why the formalism of $L$-functionals is appropriate for the consideration of scattering of thermal quasiparticles, however, this formalism can be useful also in quantum field theory, especially in  cases when unitary S-matrix does not exist.

The formalism of $L$-functionals has a limit as $\hbar \to 0.$ Therefore this formalism is very convenient in the analysis of the relation of quantum scattering with  the non-linear classical scattering and , in particular, with  the scattering of solitons\cite {CL}. 

Let us formulate some results of the paper. We  will work in the formalism of $L$-functionals representing a density matrix $K$ in the representation space of CCR by the functional (\ref {l}).  Let us consider a formal translation-invariant  Hamiltonian (\ref {h}). We denote by $S$ the renormalized scattering matrix  corresponding to (\ref {h}) regarded as an operator in Fock space. ( We assume that the ultraviolet divergences are absent; talking about the renormalization we have in mind the renormalization of one-particle energy (mass renormalization in relativistic theory) and the wave function renormalization). The scattering matrix $\hat S$ in the formalism  of $L$-functionals
is related with $S$ by the formula $\hat SL_K=L_{SKS^*}.$ We can consider this formula as a definition of $\hat S$;  it is equivalent  to the definition (\ref {SSS}) in terms of adiabatic S-matrix.
For every translation-invariant  $L$-functional  $L$
we can define the corresponding generalized Green functions (GGreen functions)  as $\bra BA \ket_L $ ( the expectation value of $BA$ in the state $L$ ) where $A$ is a chronological product of Heisenberg operators (time decreasing) and $B$ is an antichronological product of Heisenberg operators (time increasing). 
If $L$ corresponds to a density matrix $K$ then the GGreen function is equal to $Tr  BAK.$
(These functions appeared  also in Keldysh and TFD formalisms ) 

Inclusive scattering matrix can be calculated as  on-shell  GGreen function.

This is discussed in the last section that is completely independent of preceding sections ( only the definition of $L$-functional is used).

Notice that there exists an obvious generalization of all results of  present paper to the case of fermions. In this case $L$-functionals   are positive functionals  on the algebra with generators satisfying canonical anticommutation relations (Clifford algebra), $\alpha$ and $\alpha^*$ in the definition of $L$-functional are anticommuting variables.
\section{L-functionals}
Let us consider a representation of canonical commutation relations (CCR) in Hilbert space $\cal H. $ Here we understand CCR as relations 
$$[a_k,a^+_l]=\hbar \delta_{kl}, [a_k,a_l]=[a^+_k,a^+_l]=0,$$
where $k,l$ run over a discrete set $M.$
To a density matrix $K$  (or more generally to any trace class operator in $\cal H$)  we can assign a functional $L_K(\alpha^*,\alpha)$ defined by the formula
\begin{equation}
\label{l}
L_K(\alpha^*,\alpha)=Tr  e^{-\alpha a^+}e^{\alpha^*a}K
\end{equation}
Here $\alpha a^+$ stands for $\sum \alpha_k a^+_k$ and $\alpha^*a$ for $\sum \alpha^*_ka_k,$ where $k$ runs over $M.$
This formula makes sense if $\alpha\alpha^*=\sum |\alpha_k|^2<\infty.$ ( This follows from the remark that $e^{-\alpha a^++\alpha^*a}=e^{\hbar \alpha \alpha ^*/2}e^{-\alpha a^+}e^{\alpha^*a}$ is a unitary operator in $\cal H $).
We can apply (\ref {l}) also in the case when $K$ is an arbitrary operator of trace class ( not necessarily a density matrix).

 One can say that $L_K$ is a generating functional of correlation functions.

 One can consider also a more general case when CCR are written in the form
$$[a(k), a^+(k')]=\hbar\delta (k,k'), [a(k), a(k')]= [a(k)^+, a(k')^+]=0,$$
$k,k'$ run over a measure space $M$.  We are using the exponential form  of CCR; in this form a representation of CCR is specified  as a collection of  unitary operators $e^{-\alpha a^++\alpha^*a}$ obeying appropriate commutation relations.  Here $\alpha(k)$ is a complex function on the measure space $M$ , the expressions of the form $\alpha ^*a, \alpha a^+$  can be written as integrals  $\int\alpha ^*(k)a(k)dk, \int\alpha(k) a^+(k)dk$ over $M.$ In the space of CCR $a(k), a^+(k')$ become Hermitian conjugate  generalized operator functions.  We always assume that  $\alpha$ is square-integrable, then the expression (\ref {l}) is well defined   .

Knowing $L_K$ we can calculate $\langle A\rangle_L= Tr AK$ where $A$ is an element of the unital associative algebra  $\cal A$ generated by  $a_k, a^+_l $ (Weyl algebra).   Hence we can consider $L_K$ as  a positive linear functional $\langle A\rangle_L$   on  the algebra  $\cal A$.  ( We consider $\cal A$ as an algebra with involution $^+$, positivity means that  $\langle A^+A\rangle_L\geq 0.$)

A  functional $L(\alpha^*,\alpha)$ satisfying the condition $ L(-\alpha^*,-\alpha)=L^*(\alpha,\alpha^*)$ and  positivity condition is called  physical $L$-functional.  We say that such a functional is normalized if $L(0,0)=1.$  Normalized physical $L$-functionals  are in one-to-one correspondence with states on $\cal A.$ We will consider also the space $\cal L$ of all functionals  $L(\alpha^*,\alpha)$ (of all $L$-functionals).  ( More precisely one can define $\cal L$ as linear span of the set of physical $L$-functionals.) Every trace class operator $K$ in a representation space of CCR specifies an element of $\cal L$ by the formula (\ref{l}).

We can define an antilinear involution $L\to \tilde L $ on the space $\cal L$ of functionals $L(\alpha^*,\alpha)$ by the formula
\begin{equation}
\label{t}
(\tilde L)(\alpha^*,\alpha)=L^*(-\alpha,-\alpha^*).
\end{equation}
Physical $L$-functionals are invariant with respect to this involution. It is easy to check that
$$\tilde L_K =L_{K^+}.$$

Every normalized vector $\Phi\in \cal H$ specifies a density matrix and hence a physical $L$-functional. Conversely, every physical $L$-functional corresponds to a vector  in some representation of CCR , given by so called GNS construction.  One can characterize this representation by the requirement that it contains a cyclic vector $\Phi$ obeying $L(A)=\bra A\Phi, \Phi \ket.$ However, an $L$-functional can be obtained from many  density matrices in many representations of CCR.   

An action of  Weyl algebra $\cal A$  on $\cal L$   can be specified by operators $$b^+(k)=\hbar c^+_1(k)-c_2(k),b(k)=c_1(k)$$ obeying CCR. Here  $ c^+_i (k)$ are multiplication operators by $\alpha_k^*,\alpha_k$ and $c_i(k)$ are  derivatives with respect to $\alpha_k^*,\alpha_k.$ This definition is prompted by relations 
\begin{equation}
\label{b}
L_{a(k)K}=b(k)L_K, L_{a^+(k)K}=b^+(k)L_K, \end{equation}
Applying the involution $L\to\tilde L$ we obtain another representation of $\cal A$ on $\cal L$. It is specified by the operators 
$$\tilde b^+(k)=-\hbar c_2^+(k)+c_1(k), \tilde b(k)=-c_2(k), $$
obeying CCR and satisfying
\begin{equation}
\label{bb}
 L_{Ka^+(k)}=\tilde b(k)L_K,  L_{Ka(k)}=\tilde b^+(k)L_K,
\end{equation}

%\subsection{}
More generally for any $A\in \cal A$ we have an  operator acting in $\cal L$  denoted by the same symbol and obeying $A(L_K)=L_{AK}$  (the  left multiplication by $A$ in $\cal A$ specifies the action of $A$ on linear functionals.)  We can define also an operator $\tilde A$ using the formula $\tilde A( L)=A(\tilde L)$. If $L=L_K$ then $\tilde A L=L_{K^+A^+}.$ It is easy to check that $\widetilde {A+B}=\tilde {A}+\tilde B, \widetilde {AB}=\tilde {A}\tilde B, A\tilde B=\tilde B A.$ Operators $\tilde A$ specify another action of Weyl algebra on $\cal L$, that commutes with the original action; one can say that the direct sum of two Weyl algebras acts on $\cal L$.

Notice that the space of physical $L$-functionals is not invariant with respect to the operators $A$ and $\tilde A$, however, it is invariant with respect to the operators $\tilde A  A$ where $A\in \cal A.$ It is easy to check that
$$\tilde A AL_K=L_{AKA^+}.$$

Let us consider a  Hamiltonian $H$ in a space of representation of CCR.  We will write $H$ in  the form
\begin{equation}
\label{h}
H=\sum _{m,n}\sum _{k_i,l_j} H_{m,n}(k_1, ...k_m|l_1,...,l_n)a^+_{k_1}...a^+_{k_m}a_{l_1}...a _{l_n}
\end{equation}
There are two operators in $\cal L$ corresponding  to $H$: 
\begin{equation}
\label{h1}
H=\sum _{m,n}\sum _{k_i,l_j} H_{m,n}(k_1, ...k_m|l_1,...,l_n)b^+_{k_1}...b^+_{k_m}b_{l_1}...b _{l_n}
\end{equation}
(we denote it by the same symbol) and
\begin{equation}
\label{h1}
\tilde H=\sum _{m,n}\sum _{k_i,l_j} H_{m,n}(k_1, ...k_m|l_1,...,l_n)\tilde b^+_{k_1}...\tilde b^+_{k_m}\tilde b_{l_1}...\tilde b _{l_n}
\end{equation}

The equation of motion for the $L$-functional $L(\alpha^*,\alpha)$
has the form
\begin{equation}
\label{em}
i\hbar\frac{dL}{dt}=\hat {H} L= HL-\tilde HL
\end{equation}
(We introduced the notation $\hat H=H-\tilde H.$)
It corresponds to the equation for density matrices, this follows from the formula
$$\hat H L_K=L_{HK-K^+H^+}.$$

Notice that often the equations of motion for $L$-functionals make sense even in the situation when the equations of motion in the Fock space are ill-defined. This is related to the fact  that  vectors and density matrices from all representations  of CCR are described by $L$-functionals. This means that applying the formalism of $L$-functionals we can avoid the problems related to the existence of inequivalent representations of CCR. It is well known, in particular, that these problems arise for translation-invariant Hamiltonians,  in perturbation theory these problems appear as divergences related to infinite volume. Therefore in the standard formalism it is necessary to consider at first a Hamiltonian in finite volume $V$ (to make volume cutoff or, in another terminology, infrared cutoff ) and  to take the limit  $V\to \infty$ in physical quantities.
In the formalism of $L$-functionals we can work directly in infinite volume.

In what follows we consider (\ref{h}) as a formal expression; we assume that it is formally Hermitian.  There is no necessity to assume that (\ref {h}) specifies a self-adjoint operator in one of representations of CCR.

In momentum representation one can write a translation-invariant  Hamiltonian in the form 
\begin{equation}
\label{hh}
H=\sum _{m,n}\int H_{m,n}(k_1, ...k_m|l_1,...,l_n)a^+(k_1)...a^+(k_m)a(l_1)...a(l_n)d^mkd^nl
\end{equation}
where  $H_{m,n}(k_1, ...k_m|l_1,...,l_n)$ contains a  factor $\delta (k_1+ ...+k_m-l_1-...-l_n).$ (We assume that the arguments belong to $\mathbb{R}^D.$)  Such a Hamiltonian  
does not specify a self-adjoint operator in the Fock space if the Fock vacuum is not stationary, but the equation of motion for $L$-functionals can be solved ( at least in the framework of perturbation theory). One can say also that the ground state of formal Hamiltonian (\ref {hh}) does not belong to Fock space, but the corresponding $L$-functional is well defined. (Notice that we always assume that ultraviolet divergences are absent.) Similarly, the  equilibrium states for different temperatures belong to different Hilbert spaces, but all of them are represented by well defined $L$-functionals.

Let us make some remarks about adiabatic evolution in the formalism of $L$-functionals.
Let us take a  Hamiltonian that depends on time $t$,  but is changing adiabatically (very slowly). More formally we can assume that the Hamiltonian depends on a parameter $g$ and $g=h(at)$ where $a\to 0.$ Let us take a family of $L$-functionals $L(g)$ where $L(g)$ is a stationary state of the Hamiltonian $H(g)$, i.e.
$$ \hat H(g) L(g)=0.$$

It is obvious that $ L(h(at))$ obeys the equation of motion up to terms tending to zero as $a\to 0.$ ( Notice that a similar statement is wrong in the standard  Hilbert space formulation of quantum mechanics, because the vector corresponding to a state is defined  up to phase factor. We can say only that for a smooth family of eigenvectors  $\Psi (g)$ there exists a phase factor $C(g)$ such that $C(h(at))\Psi (h(at))$ obeys the equation of motion up to terms tending to zero as $a\to 0.$ )
\section {Translation-invariant Hamiltonians. One-particle states}
In what follows we consider translation- invariant Hamiltonians.

We say that  an $L$-functional  $\sigma$ is an excitation of a translation-invariant  $L$- functional $\omega$ if it coincides with $\omega$ at infinity. More precisely we should require  that  $\sigma (T_{\bx}\alpha^*, T_{\bx} \alpha))\to \omega (\alpha^*, \alpha)$ as $\bx\to \infty$. Here $T_{\bx}$ stands for spatial translation. 

We assume that the $L$-functional $\omega$ has cluster property. The weakest form of cluster property  is the requirement that  $ \bra A T_{\bx} B \ket _{\omega} -\bra A \ket_{\omega}\bra B\ket _{\omega}$ tends to zero as $\bx\to \infty$ 
for $A,B\in \cA.$

 If $\Phi$ is a vector corresponding to $\omega$ in GNS representation  space $\cal H$ then  the state corresponding to any vector $A\Phi$ where  $A\in \cal A$ is an excitation of $\omega$; this follows from cluster property.
 
 We will define elementary excitations (called particles if $\omega$ is a ground state and quasiparticles if $\omega$ is  a general translation-invariant stationary state ) in the following way.

A  {\it one-particle state } ( one-particle excitation of the state $\omega$ ) is a generalized $\cH$-valued function $\Phi (\bp)$ obeying  $\bP \Phi (\bp)=\bp \Phi (\bp),
H \Phi (\bp)=\varepsilon (\bp) \Phi (\bp)$. (More precisely, for some
class of test functions $f(\bp)$ we should have a linear map 
$f\rightarrow \Phi (f)$ of this class into $\cH$ obeying
$\bP \Phi (f)= \Phi (\bp f), H\Phi (f)=\Phi (\varepsilon (\bp) f)$. For definiteness we can assume that test functions belong to the Schwartz space ${\cal {S}}(\bR ^D)$.) { \it We require  that there exist elements $B\in \cal A$ such that } $B\Phi=\Phi (\phi)$ and moreover the functions $\phi$ obtained from these elements are dense in the space of test functions.

We have defined a {\it stable}  (quasi) particle state. If $
H \Phi (\bp)=\varepsilon (\bp) \Phi (\bp)$  approximately we should talk about almost stable  (quasi)particle state. The scattering of almost stable (quasi)particles can be considered if the collision time is much less than the lifetime of (quasi)particles.

\section {Quadratic Hamiltonians}
Let us consider as an example the simplest translation- invariant Hamiltonian
\begin{equation} \label {H0} 
H_0=\int \omega (k) a^+(k)a(k)dk
\end{equation}
where $k$ runs over $\mathbb {R}^D.$
It can be approximated by a  Hamiltonian of the form $\sum\omega_ka^+_ka_k$ having a finite number of degrees of freedom. The equilibrium state of the latter Hamiltonian can be represented by density matrix $\Omega (T)$  in the Fock space; it is easy to check that 

\begin{equation}
\label{om}
a_k\Omega(T)=e^{-\hbar \frac{\omega_k}{T}}\Omega(T)a_k, 
a^+_k\Omega(T)=e^{\frac{\hbar \omega_k}{T}}\Omega(T)a^+_k.
\end{equation} 
Applying (\ref {b}) and (\ref {bb}) we obtain equations for the corresponding $L$-functional; taking the limit we obtain for the $L$-functional corresponding to the equilibrium state in infinite volume
\begin{equation}
\label{ }
c_1(k)L_T=e^{-\frac{\hbar \omega(k)}{T}}(-\hbar c_2^+(k)+c_1(k))L_T
\end{equation}
hence
\begin{equation}
\label{ }
c_1(k)L_T=-n(k) c_2^+(k)L_T, c_2(k)L_T=-n(k) c_1^+(k)L_T
\end{equation}
where
\begin{equation}
\label{n}
n(k)=\frac{\hbar}{e^{\frac{\hbar\omega (k)}{T}}-1}
\end{equation}

We obtain
\begin{equation}
\label{ll}
L_T=e^{-\int\alpha^*(k)n(k)\alpha(k)dk},
\end{equation}
If we are interested in equilibrium state for given density we should replace $\omega (k)$ with  $\omega (k)-\mu$ in (\ref {n}) (here $\mu$ stands for chemical potential).

The Hamiltonian $\hat H$ governing the evolution of $L$-functionals can be written in the form 
$$ \hat H= \int \omega (k) b^+(k)b(k)dk-\int \omega(k)\tilde b^+(k)\tilde b(k)dk=
\int \hbar ( \omega (k) c^+_1(k)c^+_1(k)dk-\omega (k) c^+_2(k)c_2(k))dk.$$

It follows that  the vector generalized functions $\Phi_1(k)=c^+_1(k)\Phi, \Phi_2(k)=c^+_2(k)\Phi $ in the space of GNS representation corresponding to  (\ref {ll}) are one-(quasi)particle excitations.  (Notice that in this statement $n(k)$ in the formula (\ref {ll}) is an arbitrary function; then the  corresponding state is not an equilibrium, but still it is a stationary translation-invariant state.) 

Very similar considerations allow us to prove that the $L$-functional corresponding to an equilibrium state of general quadratic Hamiltonian $ H= a^+Ma^++a^+Na+aRa$ is Gaussian (has the form $e^{\alpha^*A\alpha^*+\alpha^*B\alpha+\alpha C\alpha}$).
\section {Perturbation theory}

Let us assume now that the the Hamiltonian $H$ is represented as  a sum of quadratic Hamiltonian $H_0$ and interaction Hamiltonian $H_{int}=gV.$ Then in the formalism of $L$-functionals one can introduce in  standard way the evolution operator $\hat U(t,t_0)$, the evolution operator in the interaction picture $\hat S (t, t_0)$  and the operator $\hat S_{a}$, the analog of adiabatic $S$-matrix .  If $\hat H_{a}$ governs the evolution of $L$-functional for the Hamiltonian $H_0+h(a t)H_{int}$ then we denote by $\hat U_{a}(t, t_0)$ the operator transforming the $L$-functional at the moment $t_0$ into the $L$-functional at the moment $t$ ( the evolution operator). The operator $\hat S_{a}(t,t_0)$ is defined by the formula $$\hat S_{a}(t,t_0)= e^{\frac {i}{\hbar}\hat H_0t}\hat U_{a}(t, t_0)e^{- \frac {i}{\hbar}t \hat H_0t_0}.$$ (We assume that $h(t)$ is a smooth function equal to $1$ in the neighborhood of $0$ and to $0$ in the neighborhood of infinity. It is increasing for negative $t$ and decreasing for positive $t$.) The operator $\hat S_{a}$ is defined as $\hat S_{a}(\infty, -\infty).$

The perturbation theory for   operators $\hat S_{a}(t,t_0)$ can be constructed in standard way: we apply the formula
$$\hat S_{a}(t,t_0)=T\exp(-\frac {i}{\hbar}h(a t)\hat H_{int}(t)).$$
(We use the notation $A(t)= e^{\frac {i}{\hbar}\hat H_0 t}A e^{-\frac {i}{\hbar}\hat H_0 t}.$)

  If we are interested only in the action of these operators on functionals represented by polynomials with smooth coefficients tending to zero at infinity (smooth functionals in the terminology of \cite {T}) the diagram techniques can be described as follows. The vertices  come from $-\frac {i}{\hbar}h(a t)\hat H_{int}$. To find the propagator we calculate  the $T$-product of two operators of the form $b_i^{(+)}(t)$ and express  it in normal form with respect to the operators $c^+_i, c_i$ (i.e. the operators $c_i$ are from the right) . The propagator  (that can be considered as $4\times 4$ matrix) is equal to the numerical part of this expression. In other words the propagator is given by the formula 
$$\langle T(b(k_1,t_1,{\sigma _1})b(k_2,t_2, {\sigma _2})\rangle_{L=1}.$$
Here $b(k,0,{\sigma })$ is  one of the operators $b^+,b, \tilde b^+, \tilde b.$

Let us define the adiabatic generalized Green  functions (GGreen functions) by the formula 
\begin{equation}
\label{GAR}
G^{a}_n(k_1,t_1,\sigma_1,...,k_n,t_n,\sigma_n)=
\langle T (b(k_1,t_1,\sigma_1) \cdots b(k_n,t_n,\sigma_n) \hat S_{a}(\infty, -\infty)) \rangle_{L=1} 
\end{equation}
As usual the perturbative expansion for these functions  can be constructed by the same rules as for adiabatic $S$-matrix, but the diagrams have $n$ external vertices.

Notice that  $ \hat S_{a}(\infty, -\infty) 1\to 1$ and $ \hat S_{a}(0, -\infty) 1\to \mathbf{ L}$ as $a \to 0.$ Here $\mathbf{ L}$ denotes the $L$-functional corresponding to the ground state of the Hamiltonian $H.$ (This follows immediately  from similar statement for the adiabatic evolution operators  $ \hat U_{a}(\infty, -\infty)$ and  $ \hat U_{a}(0, -\infty)$
and from the fact that the $L$-functional $L=1$ corresponds to the ground state of $H_0.$) We obtain that the adiabatic  GGreen function $G^{a}_n(k_1,t_1,\sigma_1,...,k_n,t_n,\sigma_n) $ tends to the  GGreen function 
\begin{equation}
\label{GR}
G_n(k_1,t_1,\sigma_1,...,k_n,t_n,\sigma_n)=
\langle T (\mathbf {b}(k_1,t_1,\sigma_1)\cdots \mathbf {b}(k_n,t_n,\sigma_n) ) \rangle_{\mathbf {L}}
\end{equation}
as $a \to 0.$ ( Here we use the notation $$\mathbf {b} (k,t,\sigma)=\hat S^{-1}(t,0)b( k,t,\sigma)\hat S(t,0)=\hat U^{-1}(t,0)b( k,0,\sigma)\hat U(t,0)$$ for the analog of Heisenberg operators.) This means that we can construct  the perturbation expansion for the  GGreen function $G_n(k_1,t_1,\sigma_1,...,k_n,t_n,\sigma_n)$ taking the limit $a\to 0$ in the diagrams for $G^{a}_n(k_1,t_1,\sigma_1,...,k_n,t_n,\sigma_n) $. The only modification of diagrams is in internal vertices: now these vertices are governed  by $-\frac {i}{\hbar}\hat H_{int}$.

Similar procedure can be applied for calculation of the action of  operators $\hat S_{a} (t,t_0)$ 
 on the space of functionals represented as a  product of a  smooth functional and Gaussian functional $\Lambda=e^{\lambda},$ where $\lambda$ is a quadratic expression in terms of $\alpha^*, \alpha.$ We assume that $\Lambda$ is translation- invariant and stationary with respect to the evolution corresponding to the Hamiltonian $H_0$, i.e. $\hat H_0\Lambda=0.$  ( Here $H_0$ is a translation-invariant quadratic Hamiltonian not necessarily of the form (\ref {H0}).)It is easy to check imposing some non-degeneracy conditions that there exist such operators  $\hat c^+_i(k), \hat c_i(k), i=1,2 $ obeying CCR, that $\Lambda$ can be characterized as a functional satisfying the conditions $\hat c_i(k)\Lambda=0, i=1,2, \Lambda(0,0)=0.$ Then the perturbative expression for the action  of $\hat S_{a}(t,t_0)$ on the space under consideration can be obtained by means of the diagram technique with propagators described in the same way as for $\Lambda=1,$ the only difference is that instead of normal form   with respect to the operators $c^+_i(k), c_i(k)$ we should consider normal form with respect to the operators $\hat c^+_i(k), \hat c_i(k), i=1,2 $.  Equivalently we can define the propagator by the formula
$$\langle T(b(k_1,t_1,{\sigma _1})b(k_2,t_2,{\sigma _2})\rangle_{\Lambda}.$$  Again we can define adiabatic GGreen functions 
\begin{equation}
\label{GARL}
G^{a}_n(k_1,t_1,\sigma_1,...,k_n,t_n,\sigma_n)_{\Lambda}=
\langle T (b(k_1,t_1,\sigma_1) \cdots b(k_n,t_n,\sigma_n) \hat S_{a}(\infty, -\infty)) \rangle_{\Lambda}
\end{equation}
corresponding to $\Lambda$ and introduce the diagram technique for their calculation. 
The GGreen functions corresponding to $\Lambda$ can  be defined either as limits of adiabatic GGreen functions as $a\to 0$ or by the formula
\begin{equation}
\label{GRL}
G_n(k_1,t_1,\sigma_1,...,k_n,t_n,\sigma_n)_{\mathbf{\Lambda}}=
\langle T (\mathbf {b}(k_1,t_1,\sigma_1)\cdots \mathbf {b}(k_n,t_n,\sigma_n) ) \rangle_{\mathbf {\Lambda}}
\end{equation}
where  $\mathbf {\Lambda}=\lim _{a\to 0}\hat U_{a}(0, -\infty)\Lambda$ denotes the stationary state of the Hamiltonian $H$ that we obtain from the stationary state $\Lambda$ of the Hamiltonian $H_0$ adiabatically switching  the interaction on.
The diagrams representing the functions ( \ref {GRL})  have $n$ external vertices, the internal vertices come from
$-\frac {i}{\hbar}\hat H_{int},$ the propagator is equal to $$\langle T(b(k_1,t_1,{\sigma _1})b(k_2,t_2, {\sigma _2})\rangle_{\Lambda}.$$ In particular, we can take $H_0$ of the form (\ref {H0}) and
 \begin{equation}
\label{L}
\Lambda=e^{-\int\alpha^*(k)n(k)\alpha(k)dk}.
\end{equation}

(All translation-invariant Gaussian functionals that are stationary with respect to the Hamiltonian (\ref {H0})have this form.)
 Then we obtain the following  formulas for the propagator
 $$\langle T(b^+(t)b(\tau)\rangle_{\Lambda}=\theta(t-\tau)e^{i\hbar\omega(k)(t-\tau)}n(k)+\theta(\tau-t)e^{i\hbar\omega(k)(t-\tau)}(n(k)+\hbar)r$$
$$\langle T(b^+(t)\tilde b^+(\tau)\rangle_{\Lambda}=e^{i\hbar\omega(k)(t-\tau)}(n(k)+\hbar)$$
$$\langle T(\tilde b^+(t)\tilde b(\tau)\rangle_{\Lambda}=\langle T(b^+(t)b(\tau)\rangle_{\Lambda}$$
$$\langle T(b(t)\tilde b(\tau)\rangle_{\Lambda}=e^{i\hbar\omega(k)(t-\tau)}n(k)$$
All other entries vanish.

It follows from the above formulas that the diagrams we constructed coincide with the diagrams of  Keldysh and TFD formalisms (see \cite {CU} for review of both formalisms).

We have noticed already  that the $L$-functional corresponding to an equilibrium state of quadratic Hamiltonian is Gaussian. Assuming that the equilibrium state of the     Hamiltonian $H_0+gV$ can be obtained from the equilibrium state of $H_0$ by means of adiabatic evolution we can say that  the diagram technique we have described allows us to calculate the GGreen functions in the equilibrium state.

\section{GGreen functions}
We have constructed the diagram technique for GGreen functions.  
As in standard technique  we can express all diagrams in terms of connected diagrams; moreover, connected diagrams can be expressed in terms of 1 PI diagrams. ( One says that a diagram is one particle irreducible (1 PI) if it remains connected if we remove one of edges.  Calculating a 1 PI diagram we do not take into account the contributions of external edges.))  

The contribution of a disconnected diagram is equal to the product of the contributions of its components (up to some factor taking into account the symmetry group of the diagram). The  two- point GGreen function 
\begin{equation}
\label{GRT}
G_2(k_1,t_1,\sigma_1,k_2,t_2,\sigma_2)_{\mathbf{\Lambda}}=
\langle T (\mathbf {b}(k_1,t_1,\sigma_1)\mathbf {b}(k_2,t_2,\sigma_2) ) \rangle_{\mathbf {\Lambda}}
\end{equation}
obeys the Dyson equation

\begin{equation}
\label{GRD}\begin{split}
&G_2(k_1,t_1,\sigma_1,k_2,t_2,\sigma_2)_{\mathbf {\Lambda}}\\=
&G_2(k_1,t_1,\sigma_1,k_2,t_2,\sigma_2)_{\Lambda}+\\&\int dk_2'dt_2'd\sigma'_2 dk_2''dt_2''d\sigma''_2G_2(k_1,t_1,\sigma_1,k'_2,t'_2,\sigma'_2)_{\Lambda}\times\\&M(k'_2,t'_2,\sigma'_2,k''_2,t''_2,\sigma''_2)_{\mathbf{\Lambda}}\times\\&G_2(k''_2,t''_2,\sigma''_2,k_2,t_2,\sigma_2)_{\mathbf {\Lambda}}\\
\end{split}
\end{equation}
connecting it with the propagator and self-energy operator (mass operator)  $M$ (the integration over discrete parameters is understood as summation). The generalized mass operator $M$ is defined as a sum of 1 PI diagrams. The Green functions and mass operator can be regarded as kernels of  integral operators  hence the Dyson equation can be represented in operator form
$$G_2^{\mathbf {\Lambda}}=G_2^{\Lambda}+G_2^{\Lambda}M^{\mathbf {\Lambda}}G_2^{\mathbf {\Lambda}}$$
or
\begin{equation}\label {DD}
(G_2^{\mathbf {\Lambda}})^{-1}=(G_2^{ {\Lambda}})^{-1}+M^{\mathbf {\Lambda}}.
\end{equation}
It is useful to write this equation in $(k, \epsilon )$   -representation . (Here $\epsilon $ stands for the energy variable.) . Due to translational invariance in this representation the operators entering Dyson equation are operators of multiplications by a matrix function of $(k, \epsilon ) .$ We identify the operators with these matrix functions. 
The (quasi)particles are related to the poles of  the matrix function $G_2^{\mathbf {\Lambda}}(k, \epsilon ).$ 
 Recall that we assume that the Hamiltonian $H$ is represented as  a sum of quadratic Hamiltonian $H_0=\int \omega (k) a^+(k)a(k)dk$ and interaction Hamiltonian $H_{int}=gV.$ The poles of the GGreen function  for $g=0$ ( of the propagator) are located at the points $\pm \omega (k)$  (we set $\hbar=1$); the dependence of these poles on $g$ can be found in the framework of the perturbation theory; the location of these poles will be denoted $\pm \omega (k |g).$ ( Notice that we cannot apply the perturbation theory directly, but we can use it to find zeros of the RHS of (\ref {DD}).
 The function $\omega (k|g)$ can be regarded as energy of (quasi)particle. Only  in the ground state one can hope that this function is real (thermal quasiparticles are in general unstable).
 \section {Adiabatic S-matrix}
 
The Dyson equation can be written  also for adiabatic  GGreen functions ; they can be used to describe the asymptotic behavior of   these functions for  $a\to 0$. As we have noticed  the adiabatic  GGreen functions tend to  GGreen functions as $a\to 0$, but they do not converge uniformly. However, the adiabatic  self-energy operator converges uniformly; this allows us to analyze the asymptotic behavior of GGreen functions. (The same is true for conventional Green functions). The reason for the uniform convergence is the fact that matrix function  $M$ in $(k,t)$ representation tends to zero as $t\to \infty.$ Conventional adiabatic Green functions were analyzed in \cite {LTS} , the same method was applied in \cite {T} to obtain the approximation for adiabatic  GGreen functions. These results were used to obtain the renormalized scattering matrix from adiabatic scattering matrix. Notice, that all these considerations are based on the assumption that the functions $\omega(k|g)$ are real , therefore rigorously they can be applied only to the scattering of particles (of elementary excitations of the ground state). 
Nevertheless they make sense as approximate formulas if the quasiparticles are almost stable ( we should assume that the collision time  is much less than the lifetime of quasiparticles and choose $a$ in such a way that  $\frac {1}{a}<<$ than the lifetime of quasiparticles, but $\frac {1}{a}>>$ than the collision time). 

The following statements were derived in \cite {T} 
from the results of \cite {SCH} and \cite {LTS} in the framework of perturbation theory.

{\it The scattering matrix in the formalism of $L$-functionals  can be defined as an operator  on the space of smooth $L$-functionals by the formula
\begin {equation} \label {SSS}
\hat S=\lim _{a\to 0} V_a \hat S_a V_a
\end {equation}
where $\hat S_a=\hat S_a(\infty, -\infty)$ stands for the adiabatic S-matrix,
$$ V_a=\exp i \int r_a (k )(c^+_1(k)c_1(k)-c^+_2(k)c_2(k))dk$$
and the function $r_a$ is chosen from the  requirement that $\hat S$ acts trivially on one-particle states. } (One can give an explicit expression for $r_a$ in terms of one-particle energies $\omega (k|g)$; namely $r_a(k)=\frac 1 a\int (\omega(k|h(\tau))-\omega (k))d\tau.$)

{\it One can prove the existence of the limit in (\ref {SSS})  in the framework of  perturbation theory. }(One should impose the condition $\omega (k_1+k_2)<\omega (k_1)+\omega (k_2).$ This condition means that one-particle spectrum does not overlap with multi-particle spectrum.)

The conventional renormalized  S-matrix was  related  in  \cite {LTS}  to the adiabatic S-matrices. (To obtain the S-matrix  one should multiply the adiabatic S-matrix in finite volume  by factors similar to $V_a$,  then   take the limit when the volume tends to infinity, then take the limit $a\to 0.$)  Using the methods of \cite {LTS} one can relate $\hat S$ to the conventional renormalized S-matrix $S$; we obtain
$\hat S L_K=L_{SKS^{-1}}.$
Using this formula one can express inclusive cross-section in terms of $\hat S$  (see  the next section).
 \section {Scattering of (quasi)particles. Inclusive cross-section}
%In this section we will show the way to analyze the excitations of equilibrium state and, more generally, the excitations of a state that is obtained by means of adiabatic evolution from any translation- invariant stationary Gaussian state in the framework of  perturbation theory. We will study the scattering of these excitations (of (quasi)particles).
The present section is  independent of the rest of the paper (we use only the definition of $L$-functional).

Let us start with the situation of quantum field theory when the standard scattering matrix $S$ is well defined as an operator acting in the Fock space of asymptotic states. (Strictly speaking we should  denote the  operators acting in this space as $a_{in}(k), a^+_{in}(k)$, but we use shorter notations $a(k),a^+(k)$.) We assume that the scattering matrix as well as M\o ller matrices $S_{-}, S_{+}$ are unitary. Considering the Fock space as a representation of CCR we assign an $L$-functional $L_K$ to a density matrix $K$ in the Fock space using the formula (\ref {l}).   We define the scattering matrix in the space of $L$-functionals  by the formula
\begin{equation}
\label{SS}
\hat SL_K=L_{SKS^*}.
\end{equation}
If the density matrix $K$ corresponds to a vector $\Psi$ we can represent the RHS of (\ref {SS}) as
$$Tr e^{-\alpha a^+}e^{\alpha ^*a}SKS^*=Tr e^{\alpha^* a}SKS^*e^{-\alpha a^+} =\bra e^{\alpha ^* a} S\Psi , e^{-\alpha^*a} S\Psi \ket =$$
$$\sum_n\int dp_1...dp_n \bra e^{\alpha ^*a}S\Psi | p_1,...p_n \ket \overline { \bra e^{-\alpha ^* a }S\Psi | p_1,...,p_n\ket}$$
where $|p_1,...,p_n\ket=\sqrt {\frac 1 {n!}} a^+(p_1)...a^+(p_n)\theta$ constitute an orthonormal basis in Fock space. (Here $\theta= |0\ket$ stands for Fock vacuum.) 
The expression $\bra a(k_1)...a(k_m) S\Psi | p_1,...,p_n \ket=\sqrt {\frac {(m+n)!}{n!}}\bra S\Psi | k_1,...,k_m,p_1,...,p_n \ket$ can be interpreted as the scattering amplitude of the process $\Psi \to (k_1,...k_m, p_1,...,p_n)$  (after dividing  by  numerical factor). We see that the LHS is expressed in terms of scattering amplitudes. In particular,  writing it in the form
$$\sum_{m,m'}\int dk_1...dk_mdk'_1...dk'_{m'}  \frac {(-1)^m}{m!m'!} \alpha(k_1)...\alpha (k_m)\alpha^*(k'_1)\alpha '_{m'}(k'_{m'})\hat \sigma _{m,m'}(k_1,...,k_m|k'_1,...,k'_{m'}|\Psi)$$
we obtain
$$\hat\sigma_{m,m'}(k_1,...k_m|k'_1,...,k'_{m'}|\Psi)=$$
$$\sum_n \int dp_1...dp_n\frac {\sqrt {(m+n)!} \sqrt {(m'+n)!} } {n!}\bra S\Psi|k_1,...k_m, p_1,...,p_n\ket \overline {\bra S \Psi | k'_1,...,k'_{m'},p_1,...,p_n\ket}.$$

If $m=m', k_i=k'_i$ this expression is proportional to the inclusive cross-section $\Psi \to k_1,...,k_m.$  If the initial state $\Psi$ has definite momentum $q$ then $\bra S\Psi|k_1,...,k_m,p_1, ...,p_n \ket
$ is a product of delta-function $\delta (\sum k_i +\sum p_j-q)$ coming from momentum conservation and a function that will be denoted $ \rho_{m}(k_1,p_1,...,p_n|\Psi).$   Now 
$$\hat\sigma_{m,m'}(k_1,...k_m|k'_1,...,k'_{m'}|\Psi)=
\tilde\sigma_{m,m'}(k_1,...k_m|k'_1,...,k'_{m'}|\Psi) \delta (k'_1+...+k'_m-(k_1+...+k_m))$$ where

$$\tilde\sigma_{m,m'}(k_1,...k_m|k'_1,...,k'_{m'}|\Psi)=$$ 
$$\sum_n \int dp_1...dp_n\frac {\sqrt {(m+n)!} \sqrt {(m'+n)!} } {n!} \rho_{m}(k_1,...k_m,p_1,...,p_n|\Psi)\overline {\rho_{m'}(k'_1,...,k'_n,p_1,...,p_n|\Psi)}$$ 
$$ \delta (k_1+...+k_m+p_1+...+p_n-q).$$The inclusive cross-section is proportional to $\tilde \sigma_{m,m}(k_1,...,k_m|k_1,...,k_m|\Psi)$ in this case.

We will call $\hat S$ inclusive S-matrix. Let us show that this matrix can be calculated in terms of GGreen functions (more precisely, in terms of on-shell values of these functions). Let us take a density matrix $K$ in the representation space of CCR.  We assume that  the momentum and energy operators (infinitesimal spatial and time translations) act on this space and $K$ is translation-invariant. This allows us to define Heisenberg operators $a^+(k,t), a(k,t)$ where $k$ is the momentum variable, and $t$ is the time variable. Then the corresponding GGreen function can be defined as
$Tr BAK.$
Here $A$ denotes chronological product $T$ of Heisenberg operators (the times are decreasing) and $B$ stands for antichtonological product $T^{opp}$ of Heisenberg operators (the times are increasing).

If  the density matrix $K$ corresponds a vector $\Phi$ the GGreen function
can be represented in the form
$$\bra  A\Phi, B^*\Phi\ket$$
where $B^*$ stands for chronological product of  Hermitian conjugate operators.

Let us consider now the case when $\Phi$ is the ground state.
Then one can obtain an expression of $\hat S$ in terms of  GGreen functions that is analogous to LSZ formula. It will be derived from some identities that were used in the proof of LSZ. Namely, we can use the identity
\begin{equation}
\label {LSZ}
\begin{split}
[...[S,a_{in}(k_1,\sigma_1)]...a_{in}(k_n,\sigma_n)]=\\
(-1)^n S_{-}S_{+}^*\int dt_1...dt_n L_n...L_1 T(a(k_1,t_1,\sigma_1)...a(k_n, t_n,\sigma_n))
\end{split}
\end {equation}
where $S_{-},S_{+}$ are M\o ller matrices,the scattering matrix $S$ is represented in terms of $in$-operators $a_{in}(k,1)=a_{in}^+(k)$, $a_{in}(k,-1)=a_{in}(k)$ , operators $a(k,t,\sigma)$ are Heisenberg operators $a^+(k,t), a(k,t),$ the operators $\int dt_i L_i$  in $(k,\epsilon)$ -representation can be interpreted as "mass shell operators."  ( To apply the operator  $\int dt_i L_i$  in $(k,\epsilon)$ -representation we should multiply by $ i\Lambda (k_i,\sigma_i)(\epsilon_i-\omega(k_i)) $
and take the limit $\epsilon_i\to \omega(k_i).$ Here $\omega(k)$ stands for the location of the pole of the two-point Green function and $\Lambda$ can be expressed in terms of the residue in this pole. Notice that in the transition to $(k,\epsilon)$-representation we are using direct Fourier transform for $\sigma =-1$ and inverse Fourier transform for $\sigma=1.$ )\\\
The identity (\ref {LSZ}) can be obtained , for example, from
(32.17) of \cite {MO} (by means of  conjugation with $S_{-}$). 

 It follows from (\ref {LSZ}) that Green function defined as vacuum expectation value of  chronological product $T(a(k_1,t_1,\sigma_1)...a(k_n, t_n,\sigma_n))$ is related to scattering amplitude: one should take the Fourier transform with respect to time variables and "go to mass shell" in the sense explained above. This gives the LSZ formula.  We remark that these considerations go through also in the case when instead of vacuum expectation  value $\bra 0|A|0\ket$ where $|0\ket$  obeys $a_{in} (k)|0\ket=0$ (represents physical vacuum)  we take matrix elements $\bra 0|A|p_1,...,p_n\ket$ where $|p_1,...,p_n\ket=\frac{1}{\sqrt{n!}}a_{in}^+(p_1)...a_{in}^+(p_n)|0\ket.$ This remark allows us to express "on-shell" GGreen functions as sesquilinear combinations of scattering amplitudes; comparing this expression with the formula for  $\hat SL_K$ we obtain the expression of $\hat S$ in terms of GGreen functions "on-shell" (an analog of LSZ formula). Indeed, we can consider GGreen function as vacuum expectation value of chronological product  $A$ multiplied  by antichronological  product $B$ . Using the fact that $|p_1,...,p_n\ket$ constitute a generalized orthonormal basis we can say that
 \begin {equation} \label {BA}
 \bra 0|BA|0\ket=\sum_n\int dp_1...dp_n \bra 0|B|p_1,...,p_n\ket\bra p_1,...,p_n |A|0\ket.
 \end{equation}
   This representation allows us to express on- shell GGreen functions in terms of scattering amplitudes. (The antichrononological product is related to the chronological one by Hermitian conjugation.)
 
 Let us write explicit expressions obtained  this way.
 We represent $S$  in  normal form
 $$ S=\sum _{r,s} \frac {1}{r!s!}\int dp_1...dp_rdq_1...q_s \sigma_{r,s}(p_1,...,p_r|q_1,...,q_s) a_{in}^+(p_1)...a_{in}^+(p_r) a_{in}(q_1)...a_{in}(q_s).$$
We assume that all $p_i $'s are distinct and all $q_j$'s are distinct, then the coefficient functions in normal form coincide with scattering amplitudes $\bra S a_{in}^+(q_1)....a_{in}^+(q_s)\theta|a_{in}^+(p_1)...a_{in}^+(p_r)\theta \ket.$
Let us take $\sigma_i=1$ for $i\leq m$, $\sigma_i=-1$  for $i>m$ in (\ref {LSZ}). Introducing the  notation  $q_i=k_{i-m}$ we obtain    that  $$\bra p_1,...,p_n| LHS|0\ket =\frac {1}{\sqrt {n!}} \sigma_{m+n,l}(p_1,...,p_n,k_1,...,k_m|q_1,...,q_l).$$
Here $LHS$ stands  for  the LHS of  (\ref {LSZ}). Now we can apply (\ref{LSZ}) and (\ref {BA}) to identify on- shell GGreen functions  with matrix entries of the inclusive scattering matrix $\hat  S.$

Our considerations used LSZ relations that  are based on the conjecture that the theory has particle interpretation. Moreover, the very definition of $\hat S$ that we have applied requires the existence of conventional S-matrix.  However, using  (\ref {SSS})  as the definition of the inclusive scattering matrix  $\hat S$  one can prove the relation between $\hat S$ and on -shell GGreen functions analyzing diagram techniques for these objects. This proof can be applied also in the case when the theory does not have particle interpretation. Moreover, the same ideas can be applied to quasiparticles considered as elementary excitations of  translation-invariant stationary state. These excitations are related to the poles of the two-point GGreen function; we can define the inclusive scattering matrix as
on- shell GGreen function or generalizing (\ref {SS}). Of course, this definition makes sense only if the quasiparticles are (almost) stable.

In \cite {SC} we will analyze the inclusive scattering matrix in the framework of algebraic quantum theory. 
 %(We assume that the arguments $k_1,...,k_n$ do not coincide.)  
%\section {Semiclassical approximation}\

{\bf Acknowledgments} 

I would like to thank Simons Center, IFT (Sao Paulo), IHES (Bures-sur-Yvette) and Skoltech for their hospitality. I am indebted to N. Berkovits, D. Buchholz, A. Kamenev, M. Kontsevich, G. Lechner, A. Mikhailov, N.Nekrasov, M. Rangamani for useful discussions.

\end{document}